\documentclass[aps,pra,twocolumn,showpacs]{revtex4}
\usepackage{times}
\usepackage{dcolumn}
\usepackage{bm}
\usepackage{graphicx}
\usepackage{amsmath}
\usepackage{latexsym}
\usepackage{amsfonts}
\usepackage{amssymb}
\usepackage{array}
\usepackage{epsfig}
\usepackage{float}

\newcommand{\ket}[1]{\left\vert#1\right\rangle}

\newcommand{\proj}[2]{\left\vert#1\rangle\langle#2\right\vert}

\newcommand{\be}{\begin{equation}}
\newcommand{\ee}{\end{equation}}
\newcommand{\ba}{\begin{array}}
\newcommand{\ea}{\end{array}}
\newcommand{\bqa}{\begin{eqnarray}}
\newcommand{\eqa}{\end{eqnarray}}

\begin{document}
\title{Long-range surface plasmon polariton excitation at the quantum level}

\author{D. Ballester\,$^1$}
\author{M. S. Tame\,$^1$}
\email{m.tame@qub.ac.uk}
\author{C. Lee\,$^2$} 
\author{J. Lee\,$^{2,3}$}
\author{M. S. Kim\,$^1$}
\affiliation{$^1$School of Mathematics and Physics, Queen's University,~Belfast BT7 1NN, United Kingdom\\
$^2$Department of Physics, Hanyang University, Seoul 133-791, Korea \\
$^3$Quantum Photonic Science Research Center, Hanyang University, Seoul 133-791, Korea
}

\date{\today}

\begin{abstract}
We provide the quantum mechanical description of the excitation of long-range surface plasmon
polaritons (LRSPPs) on thin metallic strips. The excitation process consists of an attenuated-reflection setup, where efficient photon-to-LRSPP wavepacket-transfer is
shown to be achievable. For calculating the coupling, we derive the first quantization of LRSPPs in the polaritonic
regime. We study quantum statistics during propagation and characterize the performance of photon-to-LRSPP quantum state transfer for single-photons, photon-number states and photonic coherent superposition states.
\end{abstract}


\maketitle

\section{Introduction}
Plasmonics~\cite{Zayats} is a rapidly growing area of research based at the
nanoscale that is currently experiencing intensive studies by researchers from many areas of the
physical sciences~\cite{electro}.
Plasmonic-based nanophotonic devices using surface plasmon polaritons (SPPs) have recently started to attract much interest from the quantum optics community for their use in quantum information processing
(QIP)~\cite{plasmonQIP,Alte,Lukin1,Lukin2}. 
At present, it is essential that practical techniques are properly developed for efficiently generating and controlling plasmonic excitations at the quantum level. In order to do this, a rigorous quantum mechanical model must be included for describing how photons and different forms of SPPs interact. With a clear description and theoretical understanding of these interactions, the rapid development of novel QIP applications, using nanostructured devices based on linear and nonlinear plasmonic effects~\cite{Lukin1,nonlinplasm} will become possible.

In this work, we adapt and extend techniques recently introduced by us~\cite{TSPP}
to provide the first {\it quantum mechanical} description of the coupling between single-photons and SPPs on thin metallic strips, also known as long-range SPPs~\cite{Econ} (LRSPPs). Here, an attenuated-reflection (ATR)
setup is described, that has so far only been considered for {\it classical} LRSPP
generation~\cite{ClassLRSPP}. In order to introduce the Hamiltonian for the interaction, we derive the first quantized description of the LRSPP fields in the
polaritonic regime~\cite{Econ}. We find that high {\it quantum efficiencies} can in fact be reached
for photon-to-LRSPP wavepacket-transfer upon appropriate modification of the ATR geometry. We
comment on the extent to which the excited LRSPPs preserve quantum statistical features of the
original photons as they propagate along realistic metallic strips. We then characterize the performance of photon-to-LRSPP {\it quantum state transfer}, focusing on an informative example of the transfer of coherent superposition states~\cite{cat}. Recently, we have become aware of an experimental effort to transfer a similar type of nonclassical field into a LRSPP~\cite{Ander}. 

The benefits of exciting LRSPPs
in this configuration compared to the previously studied standard single-interface SPPs~\cite{TSPP,OttoKret} include a significant increase in the
propagation length, together with the support for both transverse magnetic (TM) and transverse electric (TE) polarization degrees of freedom, given the correct lateral width and thickness of the metallic strip~\cite{LRSPPHV}.
Therefore they have the potential to open up a wider variety of QIP applications, where this type of additional flexibility is necessary. The work
we present here provides a valuable description of the physics of photon-SPP coupling in multilayer geometries at the
quantum level and the new methods we have developed specifically for this task should be well-suited to other complex SPP excitation scenarios, such as grating~\cite{Grating} and end-fire~\cite{EF} techniques. 

The paper is structured as follows: In Section II, we provide the quantum
mechanical description of LRSPPs. We also introduce the ATR setup used for the excitation of LRSPPs with
single photons and the coupling Hamiltonian for the fields. In Section III we analyze this coupling for single modes of the system as well as for wavepackets involving single-photons and photon-number states. From this analysis we determine the transfer efficiencies of photons to LRSPPs over a range of input frequencies. In Section IV we then examine the extent to which quantum statistics of the injected
photons are preserved during transfer and propagation of the excited LRSPPs. In Section V we characterize the performance of photon-to-LRSPP quantum state transfer, providing an illuminating example of the transfer of coherent superposition states. Finally, Section VI
summarizes our main results.
\section{Excitation setup}
\begin{figure}[b]
\centerline{\psfig{figure=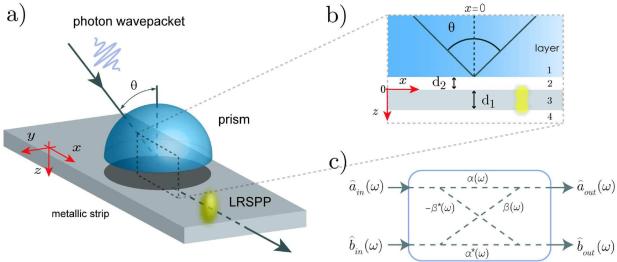,width=8.3cm,height=3.6cm}} \caption{(Color online) Single-photon excitation of LRSPPs
using attenuated-reflection. {\bf (a)}: A photon wavepacket is injected into the system at a
specific angle $\theta$, with a prism mediating an interaction between the photon and LRSPP modes.
The minimum prism size is diffraction-limited. {\bf (b)}: The ATR excitation geometry considered.
{\bf (c)}: Transfer process for the photon and LRSPP mode operators.} \label{fig:setup}
\end{figure}

\begin{figure}[t]
\centerline{\psfig{figure=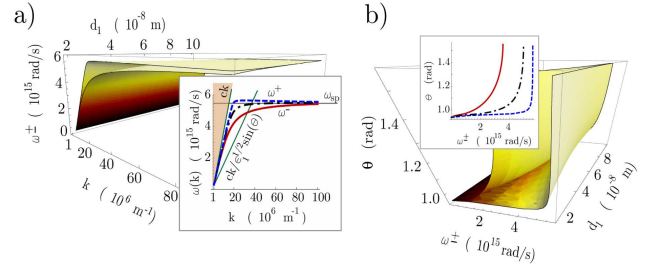,width=8.5cm,height=3.5cm}} \caption{(Color online) {\bf (a)}: Dispersion relations for the
$\omega^{\pm}$ excitations as a function of $k$ and metal thickness
$d_1$. The inset shows the curves $\omega^{+}(k)$ (dashed) and $\omega^{-}(k)$ (solid)
corresponding to $d_1=20$ $nm$, and the same curves merged at $d=100$ $nm$ (dashed-dotted). {\bf (b)}:
Coupling angle $\theta$ as a function of frequency and metal thickness $d_1$. The inset shows the coupling angle $\theta$ for $\omega^{+}(k)$ (dashed) and
$\omega^{-}(k)$ (solid) at $d_1=20$ $nm$, and the same curves merged at $d=100$ $nm$ (dashed-dotted).}
\label{fig:disp}
\end{figure}

LRSPPs are nonradiative electromagnetic excitations associated with electron charge density waves
propagating along the interfaces of a dielectric-metal-dielectric configuration~\cite{Econ}. In Figs.~\ref{fig:setup}~{\bf (a)} and~{\bf (b)} we show the ATR setup and geometry utilized for single-photon excitation of LRSPPs. It consists of four layers: a
metallic layer (with permittivity $\epsilon_3=\epsilon_m$), two
dielectric layers (with $\epsilon_2=\epsilon_4=1$), and a prism ($\epsilon_1$). We consider the
metal as silver in this work only to illustrate our main
results, with the theory developed supporting a more general setting. For LRSPP excitations, due to the collective nature of the electron
charge density waves, a macroscopic picture of the electromagnetic field produced is
appropriate~\cite{ER,Econ}. Upon quantization, LRSPPs are found to correspond to bosonic modes.
A brief
outline of this quantization is given in the Appendix. It is well-known that {\it classically} two types of TM surface modes are
found for the geometry depicted in Fig.~\ref{fig:setup}~{\bf (b)}: An antisymmetric mode, denoted by eigenfrequency $\omega^+$ and a symmetric mode denoted by
$\omega^-$. The {\it quantized} vector potential in the continuum limit for these modes propagating
along an air-metal-air interface in the $\hat{\mathbf x}$ direction, as shown in Fig.~1~{\bf (a)},
is found to be \bqa \hat{\mathbf A}_{SPP}^{\pm}({\mathbf r},t) & \propto & \int_0^{\infty}
{\mathrm d} \omega^\pm
({\cal N}^{\pm}(\omega^\pm ){\cal W})^{-1/2}\times \nonumber \\
& & [{\bm \phi}({\mathbf r},\omega )e^{-i \omega^\pm  t}\hat{b}(\omega^\pm )+ h.c]. \eqa Here ${\cal N}^{\pm}(\omega^{\pm} )$ is a
frequency dependent normalization and ${\cal W}$ is the {\it beam-width}~\cite{Blow}. For both 
$\omega^{+}$ and $\omega^{-}$ the
$\hat{b}(\omega)$'s ($\hat{b}^{\dag}(\omega)$'s) correspond to bosonic annihilation (creation) operators
which obey the 
quantum mechanical 
commutation relations
$[\hat{b}(\omega),\hat{b}^{\dag}(\omega')]=\delta(\omega-\omega')$. The modefunctions 
are given
by \bqa {\bm \phi}^\pm({\bf r},\omega^\pm )&=&e^{i {\bf k}\cdot {\bf r}}[(\hat{\bf k}-(ik/\nu_0)\hat{\bf
z})e^{\nu_0 z}
 \vartheta(-z) \nonumber \\
& & +(1-\nu_m/\epsilon_m\nu_0)[(\hat{\bf k}+(i k/\nu_m)\hat{\bf z})e^{-\nu_m z} \nonumber \\
& & \mp( \hat{\bf k}-(ik/\nu_m)\hat{\bf z})e^{\nu_m (z-d_1)}] \vartheta(z)\vartheta(d_1-z) \nonumber \\
& & \mp( \hat{\bf k}+(ik/\nu_0)\hat{\bf z})e^{-\nu_0 (z-d_1)} \vartheta(z-d_1)], \label{Phi} \eqa
where the wavevector ${\mathbf k}=k \hat{\mathbf x}$, $\vartheta(z)$ is the Heaviside step
function and the decays into the metal and air are parameterized by $\nu_m^2=k^2-\left(\omega^\pm \right)
^2 \epsilon_m /c^2$ and $\nu_0^2=k^2- \left(\omega^\pm\right) ^2/c^2$ respectively. The dispersion relation between $\omega^\pm$ and $k$ is \bqa e^{- \nu_m d_1}=\pm(\nu_m+\epsilon_m \nu_0)/(\nu_m-\epsilon_m \nu_0),
\label{dispersioneq}\eqa where $d_1$ is the thickness of the metallic strip. 
The solutions of this equation, given by $\omega^{\pm}(k)$, correspond to two
different types of coupled plasma excitations at the metal-dielectric interfaces 2/3 and 3/4 shown in Fig.~\ref{fig:setup}~{\bf (b)}, which oscillate
synchronously out-of-phase ($+$) and in-phase ($-$). The dependence of these solutions on the
wavevector and slab thickness are shown in Fig.~\ref{fig:disp}~{\bf (a)}, where silver has been chosen as an example having permittivity~\cite{JohnChrist} $\epsilon_m 
(\omega) = 1- \omega_p^2/ \omega^2  + \delta \epsilon_r,$ with $\omega_p=1.402\times 10^{16}$ rad/s and $\delta \epsilon_r= 29 \omega^2/\omega_p^2$. For a fixed value of
$d_1$, they evolve above ($+$) and beneath ($-$) the dispersion relation known for SPPs at a simple air-metal interface~\cite{Zayats}, and move closer to this curve as either $k$ or $d_1$ grows~\cite{Econ}, approaching the limiting value $\omega_{sp}$ (where $\epsilon_m=-1$) as $k \to \infty$ for any
$d_1$. While only the $\omega^+$ excitations have long-range propagation lengths,
due to damping (addressed in detail in Sec. IV), here we consider
both excitations as `long-range'
in order to give a more complete description of the physical system.

Following the diagram depicted in Fig.~\ref{fig:setup}~{\bf (a)}, let us consider an incoming photon propagating in the air with direction given by the unit vector $\hat{\mathbf k}'=\sin \theta
\hat{\mathbf x}+\cos \theta \hat{\mathbf z}$ and corresponding wavevector $\mathbf k '=
k'  \hat{\mathbf k}' =  k'_x \hat{\mathbf x}+ k'_z \hat{\mathbf z}$. The vector potential is given by~\cite{Blow} 
\begin{equation}
\hat{A}_{P}({\mathbf r},t)\propto \int_0^{\infty} {\mathrm d} \omega (\omega {\cal A})^{-1/2}[e^{i k'(\hat{\mathbf k}' \cdot {\mathbf r})} e^{-i \omega t}\hat{a}(\omega)+ h.c]. \nonumber
\end{equation} Here, ${\cal A}$ is the beam-width, with the $\hat{a}(\omega)$'s and $\hat{a}^{\dag}(\omega)$'s satisfying bosonic commutation relations. The impossibility to fulfill the mode-matching conditions between the branches
$\omega^{\pm}(k)$ and the incoming photon beam, with dispersion relation $\omega (k')=c k'_x /
\sin\theta$ (shaded region of inset in Fig.~\ref{fig:disp}~{\bf (a)}), can be overcome in the ATR configuration by using a prism with dielectric
constant $\epsilon_1>\epsilon_2=1$, placed at a distance $d_2$ over the surface of the metal. This
modifies the dispersion relation of the incoming photon beam to $\omega(k')=c k'_x / (
\sqrt{\epsilon_1} \sin\theta)$ (solid line of inset in Fig.~\ref{fig:disp}~{\bf a}). For $\theta$ greater than the critical angle, an evanescent photon field is created
below the prism surface due to total internal reflection~\cite{OttoKret}.
This provides a mechanism for achieving the coupling between the photon and LRSPP, as the $x$ component of the transmitted wavevector remains unchanged (see Fig.~\ref{fig:disp}~{\bf (b)} for mode-matching $\theta$ values). In order to model the ATR geometry, we make use of the 4-layer (4L) configuration in Fig.~\ref{fig:setup}~{\bf (b)}. The modefunctions are given by
\begin{eqnarray}
{\bf \Psi}({\mathbf r}, \omega) = r \tilde{{\bm \psi}} ({\mathbf r}, \omega) \vartheta(-(z+d_2)) + \tau
{\bm \psi}({\mathbf r}, \omega) \vartheta(z+d_2) , \label{Psi}
\end{eqnarray}
where $r$ and $\tau$ denote reflection and transmission coefficients,
$|r(\omega)|^2+|\tau(\omega)|^2=1, \forall \omega$. These coefficients, together with the modefunctions
$\tilde{\bm \psi} ({\mathbf r}, \omega)$ and ${\bm \psi} ({\mathbf r}, \omega)$ are determined by solving
Maxwell's equations for the incoming photon field, as in the {\it classical} coupled mode approach. In what follows, we will show how the $r$ and $\tau$ coefficients are combined with the quantum mechanical operators associated with the modes to derive the {\it quantum} coupling model. The modefunction $\tilde{\bm \psi} ({\mathbf r}, \omega)$ possesses a real component of the wavevector in
the $\hat{\mathbf{z}}$ direction and therefore cannot meet the mode-matching conditions required for coupling to LRSPPs. Only the modefunction ${\bm \psi}({\mathbf r}, \omega)$
is involved in the coupling of photons to LRSPP's and we have
\begin{eqnarray}
{\bm \psi}({\mathbf r}, \omega) &=& e^{i k'_x x } [(\boldsymbol{\kappa}_1 e^{-\gamma_0 (z+d_2)}
+ \boldsymbol{\kappa}_2 e^{\gamma_0 z}  ) \vartheta(-z) \nonumber \\
& & + (\boldsymbol{\kappa}_3 e^{-\gamma_m z}
+ \boldsymbol{\kappa}_4 e^{\gamma_m (z-d_1)}  ) \vartheta(z) \vartheta(d_1-z) \nonumber \\
& & + \boldsymbol{\kappa}_5 e^{-\gamma_0 (z - d_1)}  \vartheta(z-d_1) ], \label{psi}
\end{eqnarray}
where the $\boldsymbol{\kappa}_i$'s are vector-valued functions related by boundary conditions
at the interfaces, while $\gamma_{m}^2=(k'_x)^2-\epsilon_m \omega^2/c^2$ and $\gamma_{0}^2=(k'_x)^2-
\omega^2/c^2$. 

Within a linear response regime~\cite{TSPP}, the process of coupling between the photon field and the
plasmon field can be described in the Heisenberg picture by a transformation matrix $\cal{T}(\omega)$ as
\begin{equation}
\left[
\begin{array}{c}
\hat{a}_{out}(\omega) \\
\hat{b}_{out}(\omega)
\end{array}
\right] = 
\left[
\begin{array}{cc}
\alpha(\omega) & \beta(\omega) \\
-\beta^*(\omega) & \alpha^*(\omega)
\end{array}
\right]
 \left[
\begin{array}{c}
\hat{a}_{in}(\omega) \\
\hat{b}_{in}(\omega)
\end{array}
\right], \label{Heisenberg}
\end{equation}
with $|\alpha(\omega)|^2+|\beta(\omega)|^2=1, \forall \omega$. The transfer process is depicted in Fig.~\ref{fig:setup}~{\bf (c)}. The applicability of a linear
approach is fully justified in this context, as we are interested in the description of the excitation of
LRSPPs by a weak intensity photon field~\cite{linok}. The coefficients of $\cal{T}(\omega)$ are determined through
the overlap of system modefunctions, while the commutation relations of the quantum operators $\hat{a}(\omega)$ and $\hat{b}(\omega)$ properly define the structure of ${\cal T}(\omega)$ as a valid unitary quantum transfer matrix~\cite{salehleon}. The
operators $\hat{b}_{in/out}$ are associated with the in/out LRSPP modefunctions ${\bm \phi}^{\pm}({\mathbf r},\omega)$
in Eq.~(\ref{Phi}), {\it i.e.} $\hat{b}_{in}= \hat{b}$, whereas $\hat{a}_{in/out}$ are associated with the in/out modefunctions ${\bm \Psi} ({\mathbf
r},\omega)$ of the 4L configuration. Here we assume that the photon field experiences negligible losses as it enters the prism and set $\hat{a}_{in}=\hat{a}$. Thus, we have
\begin{eqnarray}
 \beta^* (\omega) &=& -\tau(\omega) \delta(\omega-\omega') \delta(k-k'_x) \int {\rm d}z  \left[\left( {\cal N}_1^{\pm}
 (\omega) \right)^{-1/2}   \right. \nonumber  \\  && \times {\bm \phi}^{\pm}(z,\omega)\Bigr]^* \cdot \left[\left({\cal N}_2(\omega') \right)^{-1/2} {\bm \psi}(z,
\omega') \right] , \label{beta*}
\end{eqnarray}
with $ \phi^{\pm}(z,\omega)$ and $\psi(z,\omega')$ denoting the $z-$dependent part of the functions in Eqs.~(\ref{Phi}) and (\ref{psi}), with normalization factors ${\cal N}_1^{\pm}$ and ${\cal N}_2$ respectively. Expression (\ref{beta*}) can be obtained using classical coupled mode theory. However, how the value of $\beta(\omega)$ enters into the quantum coupling model of Eq. (\ref{Heisenberg}) is determined by the commutation relations of the mode operators $\hat{a}(\omega)$ and $\hat{b}(\omega)$ which define ${\cal T}(\omega)$. A derivation of the coupling based solely on a classical electrodynamics approach is unsuitable when one considers properties that are explicitly dependent on the quantum operators associated with the modes of the excitations. For instance, a description of the excitation of LRSPPs by an $n$-photon state of light would not be possible. This issue is discussed in more detail in Sect. IV.

In describing the coupling process so far, we have used the Heisenberg
picture. However, in many practical situations, it is more convenient to work in the Schr\"odinger
picture, where the coupling is described by the following Hamiltonian
\begin{eqnarray}
\hat{\cal H}_S&=& \int_{0}^{\infty}{\mathrm d} \omega \hbar \omega \hat{a}^{\dag}(\omega)\hat{a}(\omega)+\int_{0}^{\infty}{\mathrm d} \omega \hbar \omega \hat{b}^{\dag}(\omega)\hat{b}(\omega)  \label{Hamil} \\
& & + i \hbar \int_{0}^{\infty}{\mathrm d} \omega [g(\omega)\hat{a}^{\dag}(\omega)\hat{b}(\omega)-g^*(\omega)\hat{b}^{\dag}(\omega)\hat{a}(\omega)], \nonumber
\end{eqnarray}
with coupling coefficient~\cite{salehleon} $g(\omega)=e^{i \arg \beta(\omega) } \sin^{-1} | \beta(\omega) |$.
\section{Photon-LRSPP transfer}
In this Section we investigate the optimization of the coupling coefficient $g(\omega)$ for a given range of
parameters $d_1$, $d_2$, and $\omega$. For clarity we will handle the dependence on all three variables by first optimizing the coupling
over $d_2$ for any pair of values of $d_1$ and $\omega$, and then optimizing over $d_1$. Here, we use a prism with $\epsilon_1=1.51$ and silver as an example, with the phenomenologically-derived~\cite{JohnChrist} dielectric function $\epsilon_m (\omega) = 1- \omega_p^2 / \left(\omega(\omega+ i \Gamma)\right) + \delta \epsilon_m$, where $\Gamma=6.25 \times 10^{13}$ rad/s and $\delta
\epsilon_m=\delta \epsilon_m^r+i \delta \epsilon_m^i$ with $\delta \epsilon_m^i=0.22$. In deriving the modefunctions ${\bm \phi}^{\pm}(\mathbf{r},\omega)$, we have assumed negligible damping in the metal for the surface plasmon field during the transfer process, allowing us to treat damping as the LRSPP propagates separately from the excitation process.

\begin{figure}[t]
\centerline{\psfig{figure=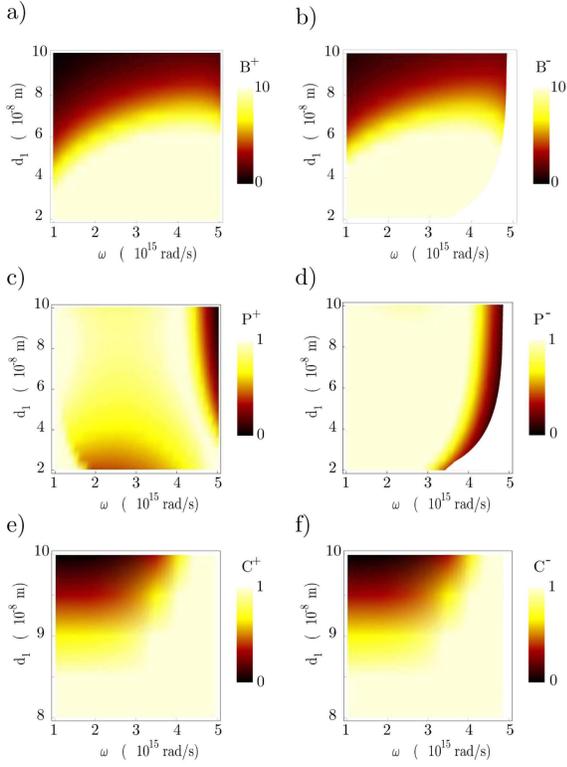,width=8.0cm,height=10.2cm}} \caption{(Color online) Values of the parameters $B^{\pm}$,
$P^{\pm}$, and $C^{\pm}$ corresponding to the optimized coupling $g^\pm(\omega)$ over the separation
$d_2$.}
\label{fig:bpc}
\end{figure}

For the optimization, some additional restrictions must be taken into account.
First, under realistic conditions, the incoming photon field does not constitute a genuine
monochromatic wave, but instead consists of a wavepacket with, for instance, a well-localized Gaussian
distribution in frequency. This
implies that both $\omega^+$ and $\omega^-$ surface plasmons are susceptible to being excited by the incoming wavepacket. For example, if the
incidence angle $\theta$ and the parameters, $d_1$, $d_2$, and $\omega$, are set in order to achieve the excitation of one of the surface plasmons, either $\omega^+$ or $\omega^-$, it is
possible that the other one might also be excited, due to
the bandwidth of the wavepacket. To limit this effect we introduce a {\it bandwidth parameter},%
$$B^\pm \equiv B(\omega^\pm) = \frac{\omega^+ - \omega^-}{2 \Delta\omega}.$$ Here, $\Delta \omega$ is the bandwidth of an incoming Gaussian wavepacket, which we set as $\Delta \omega=3.02\times 10^{13}$ rad/s, and is centered on $\omega=\omega^\pm$ for $B^{\pm}$. For a set $d_1$ and $\omega^\pm$, such that $B^\pm \geq 1$, the possibility to excite both surface plasmons
simultaneously is very low. The values of $B^{\pm}$ are shown in Figs.~\ref{fig:bpc}~{\bf
(a)} and {\bf (b)} and indicate that only large values of $d_1$  and low $\omega$ suffer from the possibility of simultaneous excitation. Note that the region corresponding to large frequencies and low $d_1$ in Fig.~\ref{fig:bpc}~{\bf (b)} has been subtracted. This is because the dispersion for the symmetric excitation $\omega^- (k)$ can never reach these frequencies for the range of $d_1$ considered (see inset of Fig.~\ref{fig:disp}~{\bf (a)}). This is applicable to all the plots for the symmetric excitation.

A second restriction related to the optimization of the coupling parameter is the extent to which the LRSPP modefunctions ${\bm \phi}^{\pm}(\mathbf{r},\omega)$ penetrate into the prism. If the LRSPP field penetrates too much, the modefunctions should be modified to include the presence of the prism. In order to check the validity of using ${\bm \phi}^{\pm}(\mathbf{r},\omega)$, we introduce a {\it penetration factor}, $P^\pm=2/\nu_0^{\pm} d_2$, which depends on the three parameters $d_1$, $d_2$, and $\omega$. Here, $P^\pm \leq 1$ signifies that ${\bm \phi}^{\pm}(\mathbf{r},\omega)$ at $z=-d_2$ is less than $2\%$ its maximum value. The dependence of
this factor on $d_1$ and $\omega$ is depicted in Figs.~\ref{fig:bpc}~{\bf (c)} and {\bf (d)}. The
values of $P^\pm$ shown correspond to the
optimized coupling coefficient $g(\omega)$ (over the parameter $d_2$) for any pair of values of $d_1$ and $\omega$. Since
the value of $g(\omega)$ could deviate significantly from the true coupling in regions where $P^\pm$ is large, due to the weak approximation of ${\bm \phi}^{\pm}(\mathbf{r},\omega)$ to the true modefunctions, we must ensure $g(\omega)$ meets the condition $P^\pm\leq 1$.
\begin{figure}[b]
\centerline{\psfig{figure=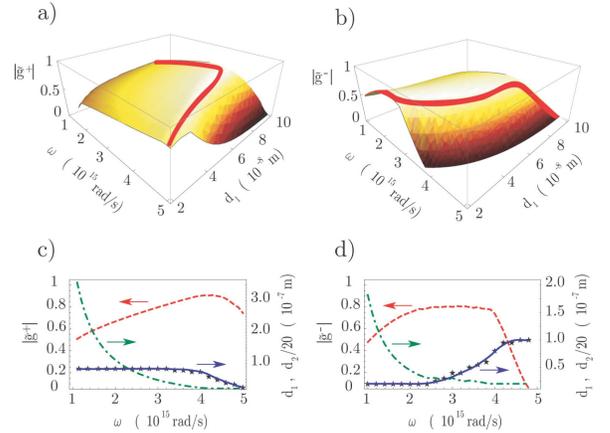,width=7.7cm}} \caption{(Color online) Optimization of the coupling over the prism separation $d_2$, subject to the constraints $B^\pm\geq 1$, $P^\pm\leq 1$, and $C^\pm\geq 1$. {\bf (a)} $\omega^{+}$ excitation, {\bf (b)} $\omega^{-}$
excitation. Panels {\bf (c)} and {\bf (d)} show the values of coupling parameter $|g^{\pm}|$ (dashed line), metal thickness $d_1$ (solid), and
prism separation $d_2$ (dot-dash), after numerical optimization: {\bf (c)}
$\omega^{+}$ excitation, {\bf (d)} $\omega^{-}$ excitation. For $d_1$, by increasing the resolution of the numerical calculation, one finds the surface plots in {\bf (a)} and {\bf (b)} become smoother and the points in {\bf (c)} and {\bf (d)} tend toward the best fit line.} \label{fig:gres}
\end{figure}

A final restriction concerns the coupled nature of the LRSPP field. LRSPPs originate from the
existence of coupled SPPs at both dielectric-metal interfaces of the metallic strip. For any finite value of the metal thickness, $d_1$,
it is always possible to find a solution of the dispersion relation Eq.~(\ref{dispersioneq}). However, the interaction
strength between both SPPs decays exponentially as $d_1$ grows and the LRSPP evolves into two single SPPs. Due to a lack of symmetry during the single-photon excitation in the
4L configuration, the incoming field may be concentrated on the nearest metal surface to the prism, without reaching the other side. If the metal thickness is large enough, then it becomes impossible
to excite any LRSPP. In order to account for this decoupling, we introduce a {\it coupled-surfaces parameter},
$C^\pm =4/\nu_m^\pm d_1$. This factor quantifies the extent to which the field penetrates into the metal (see Eq.~(\ref{Phi}))
to maintain a coupling between both SPPs, depending on the parameters $d_1$ and $\omega$. In Figs.~\ref{fig:bpc}~{\bf (e)} and {\bf (f)} the parameter $C^\pm$ remains above 1 over the entire range considered, except for large values of $d_1$ and low $\omega$. Note that the restrictions and parameters introduced here also apply to the classical case, as they depend only the classical modefunction structure.

While the penetration restriction was introduced previously in the context of single interface SPP excitation~\cite{TSPP}, the bandwidth and coupling restrictions are new parameters emerging directly from this investigation of a multilayer configuration. With all three of these important restrictions properly identified for the system, we are now in a position to correctly optimize the value of the coupling $g(\omega)$. In Figs.~\ref{fig:gres}~{\bf (a)} and {\bf (b)} we show the normalized coupling parameter,
$|\tilde{g}^\pm (\omega)|= 2 |g^\pm (\omega)| /\pi$, after being optimized over the variable $d_2$ and subject to
the restrictions $B^\pm \geq1$, $P^\pm \leq1$, and $C^\pm \geq1$. The regions where these restrictions could
not be met have been subtracted. Here a value of $|\tilde{g}^\pm|=1$ ($0$)
corresponds to the transfer of a photon field to a LRSPP with unit (zero) probability. The highlighted paths correspond to the maximum coupling achievable.
These paths are shown separately in Figs.~\ref{fig:gres}~{\bf (c)} and {\bf (d)},
plotted as a function of the frequency, together with the corresponding values of the parameters
$d_1$ and $d_2$. A maximum value of $|\tilde{g}^+| =0.9$ is achieved for the $\omega^+$ excitation, whereas the
$\omega^-$ excitation shows a flatter behavior, reaching $|\tilde{g}^- |=0.8$. It is interesting to note
that as $d_1$ increases, the optimized couplings in Figs.~\ref{fig:gres}~{\bf (a)} and {\bf (b)} move closer to those found for the photonic excitation of SPPs on
a single interface~\cite{TSPP}, but with a remarkably lower efficiency. Under such
conditions ($d_1 \to\infty$), it is only feasible to excite one of the SPPs and the excitation can no longer be considered a single LRSPP ($C^{\pm}\to 0$). This transition from multilayer coupled quantum excitation to single interface excitation should be an important factor to consider when designing optimal quantum excitation methods in the context of multiple interfaces.

\section{LRSPP propagation}
\begin{figure}[t]
\centerline{\psfig{figure=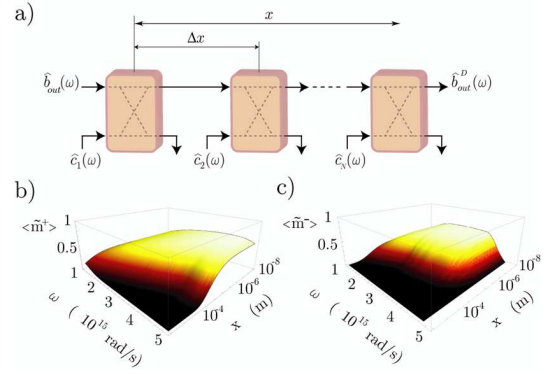,width=7.0cm}} \caption{(Color online) Phenomenological model to include
the effect of damping for the LRSPP propagation~\cite{LoudonDamp}. {\bf (a)}: Array of
beamsplitters and bath of field excitations. {\bf (b)} and {\bf (c)}: Normalized mean excitation number $\langle \tilde{m}\rangle=\langle m \rangle / n$ as a function of the frequency
and the distance traveled from the injection point. The optimal profiles obtained in Fig.~\ref{fig:gres} are used for the $\omega^{+}$ excitation {\bf (b)} and $\omega^{-}$ excitation {\bf (c)}. A
detector efficiency of $\mu=0.65$ is used~\cite{TSPP}.} \label{fig:propag}
\end{figure}

As excited LRSPPs propagate
along the metal surface they experience loss due to finite conductivity of the metal and surface roughness. This results in heating and radiative losses respectively~\cite{Raether}. For a reasonably smooth surface, thermal loss is the main source of damping~\cite{Zayats}. In order to include this loss mechanism for the LRSPP excitations, we 
follow a standard phenomenological approach, using a bath of quantized field modes~\cite{CavesCrouch,LoudonDamp,Loudon}. The main advantages of
this model are its simplicity and that it leads to the same physical
conclusions as a more rigorous derivation~\cite{Senitzky}. We consider an array of $N=x/\Delta x$ discrete,
equally spaced beamsplitters, as depicted in Fig.~\ref{fig:propag}~{\bf (a)}. The upper ports
represent the spatial evolution of the operator for a propagating LRSPP, with input
$\hat{b}_{out}(\omega)$ and output $\hat{b}_{out}^{D}(\omega)$ after a distance $x$. The lower
ports consist of a bath of field excitations, $\hat{c}_{i}(\omega)$, $i=1,\dots ,N$, which are independent and satisfy quantum mechanical bosonic commutation relations $[
\hat{c}_n(\omega),\hat{c}_m^{\dag}(\omega ')] = \delta_{nm}\delta(\omega-\omega').$ After applying
successively the beamsplitter transformations, together with the continuum limit $N\to \infty$,
$\Delta x\to 0$, such that $\hat{c}_{m}(\omega) \to \sqrt{\Delta x}\hat{c}(\omega,x')$ and
$\delta_{mn} \to \Delta x\delta(x-x')$, the operator of a damped LRSPP at point $x$ can be written as~\cite{LoudonDamp,Loudon}
\begin{equation}
\hat{b}_{out}^{D}(\omega) = e^{iKx} \hat{b}_{out}(\omega) + i \sqrt{2\kappa } \int_{0}^{x}
{\rm d}x' e^{iK(x-x')} \hat{c}(\omega,x'), \nonumber
\end{equation}
with $[\hat{c}(\omega,x), \hat{c}^{\dag}(\omega',x')] = \delta(\omega-\omega') \delta(x-x').$ Here
we have introduced the complex wavenumber $K\equiv K(\omega)=k(\omega)+ i \kappa(\omega)$. This stems from solving the dispersion relations of the LRSPP excitations with the complex-valued
dielectric function of the metal $\epsilon_m$. Under the above conditions, the output LRSPP remains a
bosonic excitation, with the appearance of the second term in $\hat{b}_{out}^{D}(\omega)$ from the
bath of field excitations preserving this bosonic nature. We now define the
time dependent creation and annihilation operators through the inverse Fourier transform of the
frequency dependent ones, for instance, $\hat{b}(t)=(2\pi)^{-1/2} \int {\rm d}\omega
e^{-i \omega t} \hat{b}(\omega) .$ Although the limits of integration over the frequency are $(-\infty, \infty)$, we are interested in the case of a narrow
wavepacket centered on frequency $\omega_0$ with bandwidth
$\Delta \omega \ll \omega_0$, thus $\omega \in (0,\infty)$ can be taken. To calculate the mean flux of LRSPPs at a point $x$ and time $t$ after their excitation, we can write~\cite{Loudon,LoudonDamp}
\begin{eqnarray}
f_{out}^{D}(t) &=& \left\langle\hat{b}_{out}^{D \dag} (t) \hat{b}_{out}^{D} (t) \right\rangle
\nonumber\\ &=& \frac{1}{2\pi} \int {\rm d}\omega \int {\rm d}\omega'
\left\langle\hat{b}_{out}^{\dag} (\omega) \hat{b}_{out} (\omega') \right\rangle  \times \nonumber\\
& & e^{-(\kappa(\omega)+\kappa(\omega')) x} e^{i[(k(\omega')-k(\omega))x-(\omega'-\omega)t]}
 ,    \label{foutD}
\end{eqnarray}
where we have used~\cite{Loudon,TSPP} $\left\langle\hat{c}^{\dag} (\omega,x) \right\rangle=
\left\langle \hat{c} (\omega,x) \right\rangle=\left\langle\hat{c}^{\dag} (\omega,x) \hat{c}
(\omega',x') \right\rangle=0$. Due to the small bandwidth of the wavepacket we are considering, the imaginary part of the LRSPP wavenumber remains essentially constant around the central
frequency, {\it i.e.} $\kappa(\omega)\approx \kappa(\omega_0)\equiv \kappa_0$, whereas the real part
can be approximated by truncating its series up to first order, $k(\omega)=k(\omega_0)+
(\omega-\omega_0)v_G^{-1}(\omega_0)$. Here $v_G^{-1}(\omega_0)=\partial k(\omega)/\partial \omega |_{\omega_0}$ is the inverse of
the group velocity of the LRSPPs at the central frequency $\omega_0$ (for either $\omega^{+}$ or $\omega^{-}$).
We then have
\begin{eqnarray}
f_{out}^{D}(t) &=&  \frac{1}{2\pi} e^{-2 \kappa_0 x} \int {\rm d}\omega \int {\rm d}\omega'
e^{i(\omega-\omega')(t-x/v_G)} \times \nonumber\\ & &  \left\langle\hat{b}_{out}^{\dag} (\omega)
\hat{b}_{out} (\omega')  \right\rangle \label{foutD2} \\ 
& = &  e^{-2 \kappa_0 x}
\left\langle\hat{b}_{out}^{\dag} (t_R) \hat{b}_{out} (t_R) \right\rangle \equiv  e^{-2 \kappa_0 x} f_{out}(t_R).  \nonumber 
\end{eqnarray}

The mean flux of LRSPPs at point $x$ and time $t$ therefore equals that at $x=0$ and time $t_R=t-x/v_G$,
but damped by a factor $e^{-2\kappa_0 x}$ due to losses incurred during the propagation. We now consider an incoming $n$-photon wavepacket state~\cite{Loudon} entering the prism with frequency profile $\xi(\omega)$, given by $\ket{n_{\xi}} =(n!)^{-1/2}( \hat{a}_{\xi}^{\dag})^{n} \ket{0}$.
Here, $\hat{a}_{\xi}^{\dag} = \int{\rm d}\omega \xi(\omega) \hat{a}^{\dag}(\omega) = \int{\rm d}t \xi(t)
\hat{a}^{\dag}(t) $ and for simplicity we use the Gaussian profile $\xi(\omega) = (2\pi \sigma^{2})^{-1/4} \exp [ -i(\omega-\omega_0)t_0 -
(\omega-\omega_0 )^{2}/4 \sigma^{2} ]$,
where $\sigma=\Delta\omega/(2\sqrt{2\ln 2})$ and $t_0$ is the time of injection. 
The mean number of LRSPPs that can be detected at point $x$ is obtained by
integrating $f_{out}^{D}(t)$ in Eq.~(\ref{foutD2}) over the time interval $\tau= [ t_0+ x/v_G - 3\sigma, t_0+ x/v_G +3\sigma
]$, leading to
\begin{equation}
\left\langle m \right\rangle = \int_\tau f_{out}^D (t) {\rm d}t = \mu e^{-2 \kappa_0 x} |\beta(\omega_0)|^{2} n  , \label{mean}
\end{equation}
where $\mu$ parameterizes the detector efficiency~\cite{LoudonLoss}, and we have used the fact the LRSPP coupling does not vary
appreciably over the bandwidth considered~\cite{TSPP}, {\it i.e.} $\beta(\omega)\approx \beta(\omega_0)$. Figs.~\ref{fig:propag}~{\bf (b)} and {\bf (c)} show $\langle \tilde{m}\rangle=\langle m \rangle / n$, for $\omega^+$ and $\omega^-$ respectively, along $x$ for the optimized $\beta(\omega)$ (and therefore $g(\omega)$) obtained from Figs.~\ref{fig:gres}~{\bf (c)} and {\bf (d)}. It is interesting to see the positive effect that the structure of the $\omega^{+}$ excitations has on damping reduction compared to $\omega^{-}$.

While the observables $\langle m^\pm \rangle$ match well the behavior of their classical counterparts~\cite{Zayats}, the field intensity $I$, we must emphasize that the formalism presented here provides a more complete description of the LRSPPs; we are now able to investigate the behavior of quantum statistics. In particular, in order to show that the LRSPP field has quantum characteristics, we need to consider the zero time-delay second-order quantum coherence function $g^{(2)}(0)$ at a fixed position~\cite{Loudon}. This observable is defined as $g^{(2)}(0)\!=\!\langle\,\colon\!\hat{I}^{2}(t)\colon\!\rangle/\langle\,\colon\!\hat{I}(t)\colon\!\rangle^2$, where $\hat{I}$ is the intensity of the quantized field operator, $\colon\colon$ denotes normal-ordering of the quantum operators and the expectation value is taken over the {\it initial} state of the field. It has been noted recently~\cite{TSPP}, that as the photon-to-surface plasmon transfer and propagation stages constitute an array of lossy beamsplitters~\cite{LoudonLoss}, $g^{(2)}(0)$, which is equal to $\langle m(m-1) \rangle/\langle m \rangle^{2}$ for an incident $n$ photon wavepacket, remains unaffected by the entire conversion process. This is to be expected~\cite{Loudon}, because
at a beamsplitter with loss coefficient $\eta^{1/2}$, the quantum observables $\langle m\rangle \to \eta \langle m\rangle$ and $\langle m(m-1) \rangle \to \eta^2 \langle m(m-1) \rangle$. Thus, the individual losses accumulated from the transfer and damping processes cancel due to the form of $g^{(2)}(0)$. For a classical field $1 \le g^{(2)}(0) \le \infty$. Thus $g^{(2)}(0)$ for a propagated LRSPP will always lie in the classically {\it forbidden} region $g^{(2)}(0)<1$. A Hanbury-Brown Twiss type experiment~\cite{HBT} could be used to measure $g^{(2)}(0)$.
Here, single-photon detection-based techniques could be employed. One might use an additional prism to convert the LRSPP excitation back into a photon and indirectly measure the signal with avalanche photodiode detectors. A more direct approach would be to use avalanche-type plasmonic detectors~\cite{pdet} embedded within the metal surface to directly probe the surface plasmon's excitation signal. In either approaches, the detection data from many identical excitation processes would be required in order to determine the overall quantum expectation value of $g^{(2)}(0)$. This repetition technique is used frequently in quantum photonic experiments~\cite{Rempe} and could be achieved easily by using a steady rate of single photons injected into the prism at set time intervals. 
\section{Quantum State Transfer}
The ability to interconvert a quantum state between two kinds of physical system is an important requirement for QIP and is one of DiVincenzo's criteria~\cite{Divincenzo} for quantum computing and communication.
As an example of efficient quantum state transfer from photons to LRSPPs using the theory developed in the previous sections, we consider a superposition of coherent states~\cite{cat}. These states reside in a single field mode consisting of superpositions of two coherent states of equal amplitude separated in phase by $180^\circ$ and written as
\begin{eqnarray}
\ket{\psi}={\cal N}(\ket{\alpha} + e^{i \varphi}\ket{- \alpha}).
\label{scs}
\end{eqnarray}
Here $\ket{\pm\alpha}={\rm exp}(-|\alpha|^2/2)\sum_{n=1}^{\infty}[(\pm \alpha)^n/\sqrt{n !}]\ket{n}$, with the normalization ${\cal N}= [2 + 2 \exp (- 2 |\alpha|^{2}) \cos \varphi]^{-1/2}$. When $\alpha$ is real, $\ket{\psi}$ for $\varphi=0$ is orthogonal to the state with $\varphi=\pi$, regardless of the size of $\alpha$. Therefore one can use $\ket{\psi}_{\varphi=0}$ and $\ket{\psi}_{\varphi=\pi}$ as a logical basis for QIP~\cite{BVE}. For the remainder of the paper, we will focus on $\ket{\psi}_{\varphi=0}$ as our example basis state. 

Photonic superpositions of coherent states injected into the prism can be transferred to a quantum state of LRSPPs under the transformation matrix ${\cal T}(\omega)$ given in Eq.~(\ref{Heisenberg}).
The unitary transformation defined by ${\cal T(\omega)}$ acting on the input product state $\ket{\Psi}_{in}=\ket{\psi}_{a_{in}} \ket{0}_{b_{in}}$ produces the state
\begin{eqnarray}
\ket{\Psi}_{out}&=&
{\cal N} (\ket{\alpha \cos g(\omega)}_{a_{out}} \ket{-\alpha \sin g(\omega)}_{b_{out}} \nonumber \\
& & + \ket{-\alpha \cos g(\omega)}_{a_{out}} \ket{\alpha \sin g(\omega)}_{b_{out}}),
\end{eqnarray}
where the coupling coefficient $g(\omega) = \sin^{-1}|\beta(\omega)|$ is used, with phase $\Phi$ absorbed into the definition of the incoming-outgoing fields~\cite{salehleon}.
Tracing out the unobserved photon mode $a_{out}$ of the photon-LRSPP system, the final state of the LRSPP mode $b_{out}$ for a specific frequency $\omega$ is given by
\begin{eqnarray}
\hat{\rho}_{b_{out}}&=& {{\cal N'}^2} (\proj{\alpha \sin g}{\alpha \sin g} + c_{0} \proj{\alpha \sin g}{-\alpha \sin g} \nonumber \\
& & + c_{0} \proj{-\alpha \sin g}{\alpha \sin g} + \proj{-\alpha \sin g}{-\alpha \sin g}), \nonumber \\
\label{density}
\end{eqnarray}
where $c_{0}=\exp[-2 \alpha^2 \cos^2 g]$.
We now consider the damped propagation of these excited LRSPP coherent superposition states as they travel along the surface of the metal. This analysis complements well the study performed in the previous section on amplitude damping of the LRSPP field. In contrast to before however, we now have a {\it coherent} superposition of an LRSPP excitation and we seek to characterize the effects of loss of coherence in this state, more commonly referred to as {\it decoherence}. Using the identity for the damped LRSPP field operator $\hat{b}_{out}^{D}$ in Section IV, and applying it to 
an excited coherent superposition state wavepacket of central frequency $\omega_0$, as described by Eq.~(\ref{density}), a straightforward calculation leads to
\begin{eqnarray}
\hat{\rho}_{b}(x)&=& {\cal N'}^2(\proj{\alpha \sin g~e^{-\kappa_{0} x}}{\alpha \sin g~e^{-\kappa_{0} x}} \nonumber \\
&& +~c_{0} c(x) \proj{\alpha \sin g~e^{-\kappa_{0} x}}{-\alpha \sin g~e^{-\kappa_{0} x}} \nonumber \\
&& +~c_{0} c(x) \proj{-\alpha \sin g~e^{-\kappa_{0} x}}{\alpha \sin g~e^{-\kappa_{0} x}} \nonumber \\
&& + ~\proj{-\alpha \sin g~e^{-\kappa_{0} x}}{-\alpha \sin g~e^{-\kappa_{0} x}}),
\label{dscs}
\end{eqnarray}
where $c(x)=\exp[-2 \alpha^2 (\sin^2 g)(1-e^{-2 \kappa_{0} x})]$.
It is easily seen from Eq.~(\ref{dscs}) that as LRSPP coherent superposition states propagate along the metal surface, the initial superposition evolves into a statistical mixture of coherent states 
due to the factors $c_{0}c(x)$. 
As the coefficients of the off-diagonal elements of the density operator expressed in the coherent state basis vanish fastest, the initial mixture of the coherent state superposition tends toward a classical mixture (dephasing) at early times, eventually moving toward the vacuum state (amplitude damping) at long times. The coefficients of the off-diagonal elements also indicate that the greater the value of $\alpha$, the more quickly quantum coherences will decay through a dephasing type process of the LRSPP state~\cite{BVK}.
\begin{figure}[t]
\centerline{\psfig{figure=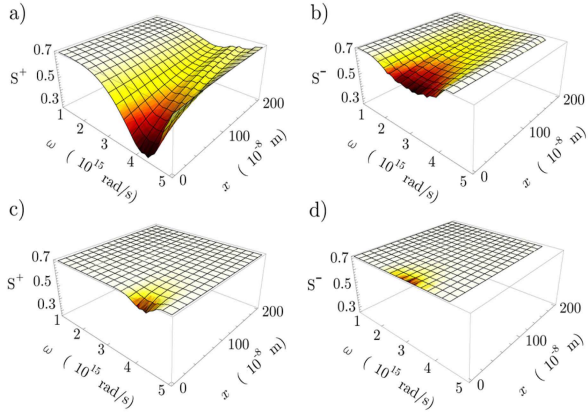,width=8.3cm,height=5.494cm}}
\caption{(Color online) The entropy ${\textrm S_{V}}$ as a function of frequency and distance traveled for fixed values of cat-state amplitude $\alpha$. {\bf (a)} and {\bf (c)} correspond to the $\omega^{+}$ LRSPP excitations with $\alpha=2$ and $5$ respectively. {\bf (b)} and {\bf (d)} correspond to the $\omega^{-}$ excitations with $\alpha=2$ and $5$ respectively.}
\label{fig7}
\end{figure}

A characterization of the loss of coherence in a quantum state can be investigated using the von Neumann entropy~\cite{von} defined as ${\textrm S_{V}}=- {\textrm Tr}(\hat{\rho}~{\rm ln} \hat{\rho})$.
The von Neumann entropy is a monotonic function of the linear entropy for a two-level quantum state of a single mode~\cite{vonlin}.
The evaluation of the von Neumann entropy requires, in general, the diagonalization of the density operator $\hat{\rho}$. 
Fortunately, the density operator of coherent superposition states with $\alpha \in {\mathbb R}$ can be decomposed into the orthonormal basis $\ket{\pm}={{\cal N_{\pm}}}(\ket{\alpha \sin g~e^{-\kappa_{0} x}} \pm \ket{-\alpha \sin g~e^{-\kappa_{0} x}})$, with ${\cal N}_{\pm}=(2\pm 2e^{-2 \alpha^2 (\sin^2 g)e^{-2 \kappa_0 x}})^{-1/2}$. In this basis, Eq.~(\ref{dscs}) becomes
\begin{eqnarray}
\hat{\rho}_{b}(x)&=& \lambda_{+}(x) \hat{\rho}_{+}(x) + \lambda_{-}(x) \hat{\rho}_{-}(x),
\end{eqnarray}
where $\lambda_{\pm}(x) = \frac{{\cal N'}^{2}}{2 {\cal N_{\pm}}^{2}}(1 \pm c_{0}c(x))$ are eigenvalues corresponding to eigenstates $\hat{\rho}_{\pm}(x) = \proj{\pm}{\pm}$, together with $\lambda_{+}(x)+\lambda_{-}(x)=1$.
The von Neumann entropy is then simply ${\textrm S_{V}(x)}=- \lambda_{+} {\textrm ln} (\lambda_{+}) - \lambda_{-} {\textrm ln} (\lambda_{-})$. In Fig.~\ref{fig7} we show the dependence of ${\textrm S_{V}}$ on the frequency and the distance traveled for specific values of $\alpha$. 
From Fig.~\ref{fig7} one can see that the greater the value of $\alpha$, the more quickly the entropy increases toward unity, and thus the mixedness of the state increases, indicating a greater loss of coherence due to dephasing, an effect noted earlier~\cite{BVK}. Moreover, one can see from the left hand ($\omega^{+}$) and right hand ($\omega^{-}$) columns of Fig.~\ref{fig7} that, for a given value of $\alpha$, the entropy slowly increases for the $\omega^{+}$ LRSPPs compared to the $\omega^{-}$ excitations as they propagate. This effect is related to the amplitude damping process observed in Sec. IV for the propagation of LRSPP-number states and is due to the smaller value of $\kappa_0$ for a given mode frequency for the $\omega^{+}$ excitations, a result of the structure of the modefunctions and their corresponding dispersion relation. According to Jeong {\it et al.}~\cite{Jeong2} mixed macroscopic superposition states can, in some cases, be more robust with respect to decoherence than their pure state counterparts. We expect this study to be useful in future work on the optimization of quantum state transfer of photons to LRSPPs and the consideration of different forms of SPPs on multiple interfaces.

\section{Concluding remarks}
We have provided a fully quantum-mechanical description of the photonic excitation of LRSPPs using a versatile ATR geometry. In order to do this it was
necessary to quantize the LRSPP field, which we included as an appendix. With this, we described the photon-LRSPP coupling mechanism by means of a linear Hamiltonian and optimized the coupling efficiency over a wide-range of parameters accessible to the setup. We found remarkably good transfer efficiencies. A
phenomenological model was then used to account for damping as the LRSPPs propagate, where the long-range behavior of the excitations manifested itself through quantum interactions with an environment. The effect of finite bandwidth for the incoming photon field on the coupling optimization and on the propagation of a LRSPP wavepacket along the metal surface was also discussed. We studied the quantum statistics of the excited LRSPP fields and provided an outline of how one might experimentally investigate them. Finally, we characterized the performance of photon-to-LRSPP quantum state transfer, providing an informative example of the transfer of coherent superposition states~\cite{cat}. We found efficient transfer and analyzed the loss of decoherence in the states. The work presented here should be a useful starting point for future research into the practical design of novel long-range and multilayer plasmonic quantum-controlled devices based at the nanoscale. Applications in this context include SPP-enhanced nonlinear photon interactions and SPP-assisted photonic quantum networking and processing.

\section{Acknowledgments}
We thank M. Paternostro and C. Di Franco for helpful discussions and insights. We acknowledge funding from ESF,
EPSRC, QIPIRC and KRF~(2005-041-C00197).
\appendix
\section*{Appendix: LRSPP quantization}
Here we develop the quantization procedure of Elson and Ritchie~\cite{ER} for the case of a thin metallic strip. An alternative approach based on the point-ion model for a dielectric slab has been used in Ref. \cite{Tomas} to quantize the long-range plasmonic fields in the polaritonic regime. For
quantization in the limit of short wavelengths, {\it i.e.} the {\it non-retarded} regime, the reader is kindly referred to Refs.~\cite{Econ,Sun}.

{\it Classical mode structure}.-- We start with the geometry depicted in Fig.~1~{\bf (b)}, where Maxwell's equations in terms of the
vector potential ${\bf A}({\bf r},t)$ lead to
\begin{gather}
\label{field1}
\left[ \nabla^2 - \frac{\epsilon}{c^2}\frac{\partial^2}{\partial t^2} \right] {\bf A}=\nabla \cdot(\nabla \cdot {\bf A}), \\
\label{field2} \nabla \cdot \dot{\bf A}=\frac{e}{\epsilon_0}(n({\bf
r},t)-n_0({\bf r})).
\end{gather}
Here $\epsilon=\epsilon({\bf r},\omega)$ is a position and frequency-dependent dielectric function, $n({\bf
r},t)$ is the electronic number density of the electron gas, $n_0({\bf r})$ is the static density
in the undisturbed electron gas and $e$ is the electronic charge. We use the gauge $\phi=0$,
where the electric field ${\bf E}=-\dot{\bf A}$ and magnetic field ${\bf B}=\nabla \times {\bf
A}$. The classical energy residing in both the fields and electron gas is given by~\cite{ER} \be \label{Ham}
{\cal H}=\frac{\epsilon_0}{2}\int {\rm d}^3r \left[\left[1+\vartheta_{eg}\frac{\omega^2(1-\epsilon)^2}{\omega_p^2({\bf r},t)}\right]\dot{\bf A}^2+ c^2 (\nabla
\times {\bf A})^2\right]. \ee 
Here, $\vartheta_{eg}=\vartheta(z)\vartheta(d-z)$ is a step function for the electron gas located in the region $0<z<d$, with $\vartheta$ denoting the Heaviside
function. In addition, $\omega_p({\bf r},t)=[n({\bf r},t)e^2/(\epsilon_0 m^*)]^{1/2}$ is a position and time dependent plasma frequency, with $m^*$ the effective electron mass. Following
a linearized hydrodynamic approach~\cite{ER,Pitarke} and taking into account the location of the electron gas, the approximation $n({\bf r},t) \approx n_0({\bf r})=\vartheta(z)\vartheta(d-z)n_0$ is used. Correspondingly, we have $\epsilon({\bf
r},\omega)=\vartheta(-z)+\vartheta(z)\vartheta(d-z)\epsilon(\omega)+\vartheta(z)\vartheta(z-d)$,
where $\epsilon(\omega)$ is a real-valued dielectric function for the metal of thickness $d$
sandwiched by two layers of air with $\epsilon=1$. Note that here we are considering an ideal case
with no damping effects in the metal. This simplifies the quantization procedure. However, damping
can be introduced at a later stage~\cite{TSPP}, as described in the main text. From
Eqs.~(\ref{field1}) and (\ref{field2}) and the above considerations, we now have the classical field equation \be
\label{finalfield} \left[ \nabla^2 - \frac{\epsilon}{c^2}\frac{\partial^2}{\partial t^2} \right] {\bf A}=0, \ee and $\nabla
\cdot{\bf A}=0$ in the region $z \neq \{0,d\}$. Here, the usual conditions of continuity of the
tangential components of the fields across the planes at $z=0$ and $d$ respectively must be
satisfied. To find the normal mode solutions for the system, we make the standard Ansatz \be
\label{Ans} {\bf A}({\bf r},t)=\sum_{\bf k}{\bf A}_{\bf k}(z)N_{\bf k}(t)e^{i {\bf k} \cdot {\bf
r}}, \ee where ${\bf r}=x\hat{\bf x} + y\hat{\bf y}$ is a vector parallel to the $x-y$ plane,
${\bf k}=k_x\hat{\bf x} + k_y\hat{\bf y}$ and the associated eigenfrequency, denoted by
$\omega_{\bf k}$, depends on ${\bf k}$. The temporal amplitude $N_{\bf k}(t)$ is assumed to
satisfy the oscillator equation, {\it i.e.} $({\rm d}^2/{\rm d}t^2+\omega_{\bf k}^2)N_{\bf
k}(t)=0$, thus upon inserting Eq.~(\ref{Ans}) into Eq.~(\ref{finalfield}), one obtains
\begin{gather}
\left(\frac{{\rm d}^2}{{\rm d}z^2}-\nu_0^2 \right){\bf A}_{\bf k}^{\pm}(z)=0,~~\left(\frac{{\rm
d}^2}{{\rm d}z^2}-\nu_m^2 \right){\bf A}_{\bf k}^{m}(z)=0, \nonumber
\end{gather}
where $\nu_m^2=k^2-\omega_{\bf k}^2\epsilon(\omega_{\bf k})/c^2$ and $\nu_0^2=k^2-\omega_{\bf
k}^2/c^2$. Here the spatial amplitudes ${\bf A}_{\bf k}^{+}(z)$, ${\bf A}_{\bf k}^{-}(z)$ and
${\bf A}_{\bf k}^{m}(z)$ correspond to fields in the $z>d$, $z<0$ and $0<z<d$ regions
respectively. Consider the following solutions: ${\bf A}_{\bf k}^{+}(z)={\bf A}_{\bf
k}^{+}e^{-\nu_0(z-d)}$, ${\bf A}_{\bf k}^{-}(z)={\bf A}_{\bf k}^{-}e^{\nu_0 z}$ and ${\bf A}_{\bf
k}^{m}(z)={\bf A}_{\bf k}^{m^+}e^{-\nu_m z}+{\bf A}_{\bf k}^{m^-}e^{\nu_m(z-d)}$, where ${\bf
A}_{\bf k}^{+}=\alpha^{+}_{\bf k}\hat{\bf k}+\beta^{+}_{{\bf k}}\hat{\bf z}$, ${\bf A}_{\bf
k}^{-}=\alpha_{\bf k}^{-}\hat{\bf k}+\beta_{{\bf k}}^{-}\hat{\bf z}$ and ${\bf A}_{\bf
k}^{m^\pm}=\alpha^{m^\pm}_{\bf k}\hat{\bf k}+\beta^{m^\pm}_{{\bf k}}\hat{\bf z}$.  
Here we focus on TM modes of the system due to boundary conditions~\cite{Econ, LRSPPHV}. By requiring that the tangential
components of ${\bf E}$ and ${\bf B}$ derived from {\bf A} across the planes $z=0$ and $d$ are
continuous and $\nabla \cdot {\bf A}=0$ elsewhere, one finds the $\alpha_{\bf k}$'s and
$\beta_{\bf k}$'s are related to one another, with solutions existing only if $e^{- \nu_m
d}=\pm(\nu_m+\epsilon(\omega_{\bf k})\nu_0)/(\nu_m-\epsilon(\omega_{\bf k})\nu_0)$ is satisfied.
At a set thickness $d$ there are two possible
eigenfrequencies of this equation for a given ${\bf k}$, which we denote $\omega_{\bf k}^\pm$.
Thus, there are two sets of coefficients for the $\alpha_{\bf k}$'s and $\beta_{\bf k}$'s for a
given ${\bf k}$. The most general form of ${\bf A}({\bf r},t)$ is then \be \label{soln} {\bf
A}^\pm({\bf r},t)=\sum_{\bf k}{\bm \phi}_{\bf k}^\pm(z)N_{\bf k}^\pm(t) e^{i {\bf k}
\cdot {\bf r}}+ c.c, \ee where the eigenmodes are given by \bqa
{\bm \phi}_{\bf k}^\pm(z)&=&[(\hat{\bf k}-(ik/\nu_0)\hat{\bf z})e^{\nu_0 z} \vartheta(-z) \nonumber \\
& & +(1-\nu_m/\epsilon(\omega_{\bf k})\nu_0)[(\hat{\bf k}+(i k/\nu_m)\hat{\bf z})e^{-\nu_m z} \nonumber \\
& & \mp( \hat{\bf k}-(ik/\nu_m)\hat{\bf z})e^{\nu_m (z-d)}] \vartheta(z)\vartheta(d-z) \nonumber \\
& & \mp( \hat{\bf k}+(ik/\nu_0)\hat{\bf z})e^{-\nu_0 (z-d)} \vartheta(z-d)], \eqa and the
time-dependent amplitudes are given by $N_{\bf k}^\pm(t)=N_{\bf k}^{\pm}e^{-i \omega^{\pm}_{\bf k} t}$. In the above, we have used $\alpha_{\bf
k}^{-}$ as the free coefficient in the coupled boundary equations and absorbed it into the definition of $N_{\bf k}^{\pm}$.
Due to the symmetry in the phases of the amplitudes in the ${\bm \phi}_{\bf k}^\pm(z)$'s with respect
to the center of the metal, the associated field modes are commonly referred~\cite{Zayats,Econ} to as antisymmetric
($\omega_{\bf k}^{+}$) and symmetric modes $(\omega_{\bf k}^{-})$.

{\it Discretization and quantization}.-- We now discretize the classical system and quantize it. The components of ${\bf k}$ are taken
to be $k_x=2 \pi \ell_x/L$ and $k_y=2 \pi \ell_y/L$, where $\ell_x$ and $\ell_y$ are integers,
with $e^{i {\bf k}\cdot {\bf r}}$ satisfying boundary conditions at the planes $x=\pm L/2$ and
$y=\pm L/2$. Substituting Eq.~(\ref{soln}) into Eq.~(\ref{Ham}) one finds the total energy of the
discretized classical modes given by $ {\cal H}^{\pm}=\sum_{\bf k}\epsilon_0 L^2 (\omega_{\bf
k}^{\pm})^2{\cal N}_{\bf k}^{\pm}(N_{\bf k}^{\pm}N_{\bf k}^{\pm *}+N_{\bf k}^{\pm *}N_{\bf
k}^{\pm}),$ where ${\cal N}_{\bf k}^{\pm}$ is a coefficient with dimensions of length. Using the
correspondence with a quantized harmonic oscillator~\cite{Loudon}, {\it i.e.} $N_{\bf
k}^{\pm}\to(\hbar/2\epsilon_0L^2\omega_{\bf k}^{\pm}{\cal N}_{\bf k}^{\pm})^{1/2}\hat{b}_{{\bf k},\pm}$
and $N_{\bf k}^{\pm *}\to(\hbar/2\epsilon_0L^2\omega_{\bf k}^{\pm}{\cal N}_{\bf
k}^{\pm})^{1/2}\hat{b}_{{\bf k},\pm}^{\dag}$, we have the Hamiltonian \be \hat{\cal
H}^{\pm}=\frac{1}{2}\hbar \omega_{\bf k}^{\pm}(\hat{b}_{{\bf k},\pm}\hat{b}_{{\bf
k},\pm}^{\dag}+\hat{b}_{{\bf k},\pm}^{\dag}\hat{b}_{{\bf k},\pm}) , \ee along with the vector
potential converted to the operator \be \hat{\bf A}^{\pm}({\bf r},t)=\sum_{\bf
k}(\hbar/2\epsilon_0L^2\omega_{\bf k}^{\pm}{\cal N}_{\bf k}^{\pm})^{1/2}[{\bm \phi}_{\bf k}^\pm({\bf
r})e^{-i \omega_{\bf k}^{\pm}t}\hat{b}_{{\bf k},\pm}+ h.c]. \nonumber \ee Here we have
${\bm \phi}_{\bf k}^\pm({\bf r})={\bm \phi}_{\bf k}^\pm(z)e^{i{\bf k}\cdot {\bf r}}$, where the creation
and annihilation operators for the quantum excitations satisfy $[\hat{b}_{{\bf
k},\pm},\hat{b}_{{\bf k}',\pm}^{\dag}]=\delta_{{\bf k}, {\bf k}'}$.

{\it Continuum limit and beam-width}.--
We now take the continuum limit using the transformations $\sum_{\bf k} \to (L/2\pi)^2\int{\rm
d}{\bf k}$ and $\hat{b}_{{\bf k},\pm}\to (2 \pi/L)\hat{b}_{\pm}({\bf k})$, leading to \bqa
\hat{\bf A}^{\pm}({\bf r},t)&=&\frac{1}{2\pi}\int {\rm d}{\bf
k}(\hbar/2\epsilon_0\omega^{\pm}(k){\cal N}^{\pm}(k))^{1/2}\times
\nonumber \\
& & [{\bm \phi}^\pm({\bf r}, {\bf k}) e^{-i \omega^{\pm}(k)t} \hat{b}_{\pm}({\bf k})+ h.c]. \eqa Next,
the excitations propagating in the $\hat{\bf x}$ direction have a beam-width
${\cal W}$ imposed in the $\hat{\bf y}$ direction~\cite{Blow} using $\int {\rm d}{\bf
k} \to (2 \pi/{\cal W})\sum_{k_y}\int {\rm d}k_x$ and setting $k_y=0$, with $\delta^2({\bf k}-{\bf
k}')\to ({\cal W}/2\pi)\delta(k-k')$ and $\hat{b}_{\pm}({\bf k}) \to ({\cal
W}^{1/2}/2\pi)\hat{b}_{\pm}(k)$, giving \bqa \hat{\bf A}^{\pm}({\bf r},t)&=&\frac{1}{2\pi}\int
{\rm d}k(\hbar/2\epsilon_0{\cal W}\omega^{\pm}(k){\cal N}^{\pm}(k))^{1/2}\times
\nonumber \\
& & [{\bm \phi}^\pm({\bf r},k) e^{-i \omega^{\pm}(k)t} \hat{b}_{\pm}(k)+ h.c]. \eqa Finally, we
convert to the frequency domain using ${\rm d}k \to [v_G(\omega^\pm)]^{-1}{\rm
d}\omega^\pm$ and $\hat{b}_{\pm}(k)\to [v_G(\omega^\pm)]^{1/2}\hat{b}_{\pm}(\omega^\pm)$, where
$v_G(\omega^\pm)=\partial \omega^\pm/\partial k$ is the group velocity. This gives the quantized
vector potential for the $\omega^{\pm}$ field as 
\bqa \hat{\bf A}^{\pm}({\bf r},t)&=&\frac{1}{2\pi}\int {\rm
d}\omega^{\pm}(\hbar/2\epsilon_0{\cal W}\omega^{\pm}v_G(\omega^\pm){\cal N}^\pm(\omega^\pm))^{1/2}\times
\nonumber \\
& & [{\bm \phi}^{\pm}({\bf r},\omega^\pm) e^{-i \omega^{\pm}t} \hat{b}_\pm(\omega^\pm)+ h.c]. \eqa



\begin{thebibliography}{99}

\bibitem{Zayats} A. V. Zayats, I. I. Smolyaninov and A. A. Maradudin, Phys. Rep. {\bf 408}, 131 (2005).

\bibitem{electro} W. L. Barnes, A. Dereux and T. W. Ebbesen, Nature {\bf 424}, 824 (2003).

\bibitem{plasmonQIP} J. L. van Velsen, J. Tworzydlo and C. W. J. Beenakker, Phys. Rev. A {\bf 68} 043807 (2003); S. Fasel, M. Halder, N. Gisin and H. Zbinden, New J. Phys. {\bf 8}, 13 (2006); X.-F. Ren, G.-P. Guo, Y.-F. Huang, Z.-W. Wang, and G.-C. Guo, Opt. Lett. {\bf 31}, 2792 (2006); A. Kamli, S. A. Moiseev and B. C. Sanders, Phys. Rev. Lett. {\bf 101}, 263601 (2008).

\bibitem{Alte} E. Altewischer, M. P. van Exter and J. P. Woerdman, Nature {\bf 418}, 304 (2002); E. Moreno, F. J. García-Vidal, D. Erni, J. I. Cirac and L. Martín-Moreno, Phys. Rev. Lett. {\bf 92}, 236801 (2004); S. Fasel, F. Robin, E. Moreno, D. Erni, N. Gisin, and H. Zbinden, Phys. Rev. Lett. {\bf 94}, 110501 (2005); X.-F. Ren, G. P. Guo, Y. F. Huang, C. F. Li and G. C. Guo, Europhys. Lett. {\bf 76}, 753 (2006).

\bibitem{Lukin1} D. E. Chang, A. S. S\o rensen, P. R. Hemmer and M. D. Lukin, Phys. Rev. Lett. {\bf 97}, 053002 (2006); A. V. Akimov, A. Mukherjee, C. L. Yu, D. E. Chang, A. S. Zibrov, P. R. Hemmer, H. Park and M. D. Lukin, Nature {\bf 450}, 402 (2007).

\bibitem{Lukin2} D. E. Chang. A. S. S\o rensen, E. A. Demler, M. D. Lukin, Nature Phys. {\bf 3}, 807 (2007).

\bibitem{nonlinplasm}
G. I. Stegeman, J. I. Burk and D. G. Hall, Appl. Phys. Lett. {\bf 41}, 906 (1982); R. T. Deck and
D. Sarid, J. Opt. Soc. Am. {\bf 72}, 1613 (1982); J. C. Quail, J. G. Rako, H. J. Simon and R. T.
Deck, Phys. Rev. Lett. {\bf 50}, 1987 (1983); I. I. Smolyaninov, A. V. Zayats, A. Gungor and C. C.
Davis, Phys. Rev. Lett. {\bf 88}, 187402 (2002); M. D. Lukin and A. Imamoglu, Nature {\bf 413},
273 (2001).

\bibitem{TSPP} M. S. Tame, C. Lee, J. Lee, D. Ballester, M. Paternostro, A. V. Zayats and M. S. Kim, Phys. Rev. Lett. {\bf 101}, 190504 (2008).

\bibitem{Econ} E. N. Economou, Phys. Rev. {\bf 182}, 539 (1969).

\bibitem{ClassLRSPP} A. Otto, Z. Phys. {\bf 219}, 227 (1969).

\bibitem{cat} Letter from Einstein to Schr\"odinger of 22 December 1950, in {\sl Briefe zur Wellenmechanik}, edited by K. Przibram (Springer, Vienna, 1963), p. 36; E. Schr\"odinger, Naturwissenschaften {\bf 23}, 807 (1935); B. Yurke and D. Stoler, Phys. Rev. Lett. {\bf 57}, 13 (1986).

\bibitem{Ander} A. Huck, S. Smolka, P. Lodahl, A. S. S\o rensen, A. Boltasseva, J. Janousek and U. L. Andersen, arXiv:0901.3969 (2009).

\bibitem{OttoKret} A. Otto, Z. Phys. {\bf 216}, 398 (1968); E. Kretschmann and H. Raether, Z. Naturforsch {\bf 23a}, 2135 (1968); E. Kretschmann, Z. Phys. {\bf 241}, 313 (1971).

\bibitem{LRSPPHV} P. Berini, Phys. Rev. B {\bf 61}, 10484 (2000); P. Berini, {\it ibid.} {\bf 63}, 125417 (2001).

\bibitem{Grating} Y. Teng and E. A. Stern, Phys. Rev. Lett. {\bf 19}, 511 (1967).

\bibitem{EF} H. P. Hsu, A. F. Milton and W. K. Burns, Appl. Phys. Lett. {\bf 33}, 603 (1978); G. I. Stegeman, R. F. Wallis and A. A. Maradudin, Opt. Lett. {\bf 8}, 386 (1983).

\bibitem{ER} J. M. Elson and R. H. Ritchie,\,Phys.\,Rev.\,B\,{\bf 4},\,4129\,(1971); J. Nkoma, R. Loudon and D. R. Tilley, J. Phys. C: Solid State Phys. {\bf 7}, 3547 (1974); M. S. Toma$\check{\rm s}$ and M. $\check{\rm S}$unji$\acute{\rm c}$, Phys. Rev. B {\bf 12} 5363 (1975); Y. O. Nakamura, Prog. Theor. Phys. {\bf 70}, 908 (1983).

\bibitem{Blow} K. J. Blow, R. Loudon, S. Phoenix and T. J. Shepherd, Phys. Rev. A {\bf 42}, 4102 (1990).

\bibitem{linok}
F. Brown, R. E. Parks and A. M. Sleeper, Phys. Rev. Lett. {\bf 14} 1029 (1965); H. J. Simon, D. E.
Mitchell and J. G. Watson, Phys. Rev. Lett. {\bf 33}, 1531 (1974); C. K. Chen, A. R. B. de Castro
and Y. R. Shen, Phys. Rev. Lett. {\bf 46}, 145 (1981); T. Y. F. Tsang, Opt. Lett. {\bf 21}, 245
(1996).

\bibitem{JohnChrist} H. Ehrenreich and H. R. Philipp, Phys. Rev. {\bf 128}, 1622 (1962); 
P. B. Johnson and R. W. Christy, Phys. Rev. B {\bf 6}, 4370 (1972).

\bibitem{salehleon} R. A. Campos, B. E. A. Saleh and M. C. Teich, Phys. Rev. A {\bf 40}, 1371 (1989); U. Leonhardt, Rep. Prog. Phys. {\bf 66}, 1207 (2003).

\bibitem{Loudon} R. Loudon, {\sl The Quantum Theory of Light}, 3$^{\rm rd}$ Ed., Oxford University Press, Oxford (2000).

\bibitem{Raether} H. Raether, {\sl Surface Plasmons}, Springer-Verlag, Berlin (1986).

\bibitem{CavesCrouch} C. M. Caves and D. D. Crouch, J. Opt. Soc. Am. B {\bf 4}, 1535 (1987).

\bibitem{LoudonDamp} J. Jeffers, N. Imoto and R. Loudon, Phys. Rev. A {\bf 47}, 3346 (1993).

\bibitem{Senitzky} I. R. Senitzky, Phys. Rev. {\bf 119}, 670 (1960).

\bibitem{LoudonLoss} H. P. Yuen and J. H. Shapiro, IEEE Trans. Inf. Theor. {\bf 26}, 78 (1980).

\bibitem{HBT} R.\,Hanbury-Brown\,and\,R.\,Q.\,Twiss,\,Nature\,{\bf 177},\,27\,(1956).

\bibitem{pdet} H. Ditlbacher, F. R. Aussenegg, J. R. Krenn, B. Lamprecht, G. Jakopic and G. Leising, App. Phys. Lett. {\bf 89}, 161101 (2006).  

\bibitem{Rempe} A. Kuhn, M. Hennrich and G. Rempe, Phys. Rev. Lett. {\bf 89}, 067901 (2002); M. Hennrich, T. Legero, A. Kuhn and G. Rempe, New J. Phys. {\bf 6}, 86 (2004).

\bibitem{Sun} M. $\check{\rm S}$unji$\acute{\rm c}$ and A. A. Lucas, Phys. Rev. B {\bf 3}, 719 (1971).

\bibitem{Pitarke} J. M. Pitarke, V. M. Silkin, E. V. Chulkov and P. M. Echenique, Rep. Prog. Phys. {\bf 70}, 1 (2007).

\bibitem{Divincenzo} D. P. DiVincenzo, Fortschr. Phys. {\bf 48}, 771 (2000).

\bibitem{BVE} S. Braunstein and P. van Loock, Rev. Mod. Phys. {\bf 77}, 513 (2005).

\bibitem{von} J. von Neumann, {\sl Mathematical Foundations of Quantum Mechanics}, Princeton University Press, Princeton (2000).

\bibitem{vonlin} H. Moya-Cessa, Physics Reports {\bf 432}, 1 (2006).

\bibitem{BVK} V. Bu$\check{\rm z}$ek, A. Vidiella-Barranco and P. L. Knight, Phys. Rev. A {\bf 45}, 6570 (1992).

\bibitem{Jeong2} H.\,Jeong, J. Lee and H. Nha, J. Opt. Soc. Am. B {\bf 25}, 1025 (2008).

\bibitem{Tomas} M. S. Toma$\check{\rm s}$, M. $\check{\rm S}$unji$\acute{\rm c}$, and Z. Lenac, Fizika {\bf 14}, 77 (1982); Z. Lenac and M. S. Toma$\check{\rm s}$, J. Phys. C: Solid State Phys. {\bf 16}, 4273 (1983).

\end{thebibliography}
\end{document}